\documentclass[twocolumn,showpacs,preprintnumbers,amsmath,amssymb]{revtex4}
\usepackage{graphicx}% Include figure files
\usepackage{dcolumn}% Align table columns on decimal point
\usepackage{bm}% bold math

\begin{document}

%\preprint{APS/123-QED}

\title{On the zero-resistance states generated by radiation in GaAs/AlGaAs
}% Force line breaks with \\

\author{S. Fujita}
 \email{Fujita@buffalo.edu}
%\altaffiliation[Also at ]{Physics Department, XYZ University.}%Lines break automatically or can be forced with \\
\author{K. Ito}%
 \email{Keiito@buffalo.edu}
 \altaffiliation[Also at ]{Research Division, National Center for University Entrance Examinations,
 2-19-23 Komaba, Meguro-ku, Tokyo 153-8501, Japan}%Lines break automatically or can be forced with \\
 \email{ito@rd.dnc.ac.jp}
\affiliation{%
University at Buffalo, SUNY, Buffalo, NY 14260, USA%\\
%This line break forced with \textbackslash\textbackslash
}%

%\author{Charlie Author}
% \homepage{http://www.Second.institution.edu/~Charlie.Author}
%\affiliation{
%Second institution and/or address\\
%This line break forced% with \\
%}%

\date{\today}% It is always \today, today,
             %  but any date may be explicitly specified

\begin{abstract}
The applied radiation excites ``holes". The condensed composite (c)-bosons formed in the excited channel
create a superconducting state with an energy gap. The supercondensate suppresses the non-condensed c-bosons
at low temperatures, but it cannot suppress the c-fermions in the base channel, and the small normal current
accompanied by the Hall field yields a $B$-linear Hall resistivity.
\end{abstract}

% PACS, the Physics and Astronomy
                             % Classification Scheme.
%\keywords{Suggested keywords}%Use showkeys class option if keyword
                              %display desired
\maketitle

Recently Mani \textit{et al}. [1] observed a set of zero-resistance (superconducting) states in GaAs/AlGaAs
heterojunction subjected to radiations at the low temperatures ($\sim 1.5$ K) and the magnetic fields
($\sim 0.2$ T). The (diagonal) resistance $R_{xx} \equiv R$ rises exponentially and symmetrically on both
sides of the fields centered at $B=4/5\,B_{f},$ $4/9\,B_{f}$, $B_{f}=\omega m^{*}/e$,
$\omega=\text{radiation frequency}$, $m^{*}=\text{effective mass}$, $e=\text{electron charge}$, indicating an
energy gap $\epsilon_{g}$ in the elementary excitation spectrum.
The phenomenon is similar to that observed in the same system in the traditional quantum Hall effect (QHE) regime
($T \sim 0.5$ K, $B \sim 10$ T) [2], with the main difference that the superconducting states are not
accompanied by the Hall resistivity plateaus for the system subjected to radiation. We call the temperature below
which the superconducting state appears the critical temperature $T_{c}$. The $T_{c}$ (1.5 K) at $B=4/5\,B_{f}$
is considerably higher than the traditional QHE critical temperature (0.5 K). Zudov \textit{et al}. [3] reported
similar magnetotransport properties for the system subjected to radiation with slightly different experimental
conditions [3]. They suggested that the principal resistivity minima occur at $B=2/(2j+1)B_{f}$,
$j=1,\,2,\,\cdots$, rather than $B=4/(4j+1)B_{f}$ (Mani's case). They also noted a noticeable side resistivity
minimum besides the principal set of the minima.

In finer experiments Mani \textit{et al}. [4] observed (a) that the deviation in the Hall resistance
$\Delta R_{H} \equiv R_{H} - R_{H, \text{dark}}$, correlates with the resistance $R$ such that $\Delta R_{H}$
nearly vanishes when $R=0$, and (b) that $\Delta R_{H}$ is negative, and it is antisymmetric with respect to small
$B$-fields: $\Delta R_{H} = \alpha B$, $\alpha=\text{constant}$, for small $B$.
The second property means that there is a current
due to ``hole"-like particles having the charge of the opposite sign to that of the main current carrier charge.
We suspect that the applied radiation generates the ``holes". This may be checked by applying the
circularly polarized lasers. The small slope $d\Delta R_{H}/dB=\alpha$ means that the ``hole"-like particle
density is considerable higher than the ``electron"-like particle density. Mani \textit{et al}. [1] suggested for
the cause of the spectral gap the electron pairing due to the excitions induced by radiation. Other mechanisms
were proposed by several theoretical groups [5]. But none of them are conclusive. In particular no explanation is
given to the question why the superconducting state can occur without the Hall resistivity plateau.

Earlier Fujita \textit{et al}. [6] developed a microscopic theory of the QHE
based on the electron(fluxon)-phonon interaction. In this theory the composite (c-) particles (bosons or fermions)
having a conduction electron and a number of flux quanta (fluxons) are bound by the phonon-exchange attraction.
The composites move as bosons (fermions) depending on the odd (even) number of fluxons in them. At the Landau
level (LL) occupation number (filling factor) $\nu=1/Q$, $Q$ odd, the c-bosons with $Q$ fluxons are generated,
and condense below the critical temperature $T_{c}$. The Hall resistivity plateau is due
to the Meissner effect, see below.

In the present work we shall extend our theory to the excited channel and show that the main series represent
the integer QHE with the $\nu=1$ state corresponding to the superconducting state $B=4/5\,B_{f}$ in Mani's case
(the state $B=2/3\,B_{f}$ in Zudov's case). The state in the base channel at $\nu=1$ remains fermionic,
accounting for the small current and the $B$-linear behavior in the Hall resistance.  If we write $\nu=P/Q$, the
Mani-series 4/(4j+1) can be recovered by setting $j=P$, $Q=1$. We predict that the fractional QHE should exist for
$Q=3,\,5,\,\cdots$. The aformentioned side dip observed by Zudov \textit{et al}. [3] should correspond to the
state at $\nu=4/3$. This dip is missing in Mani \textit{et al}. experiments. We believe that this is because the
experimental temperature $T=0.5$ is so low that the $\nu=1$ state overshadowed it. All of the superconducting
states $\nu=P/Q$, $P=2,\,3,\,\cdots$, disappear at 0 K.

GaAs forms a zinc blende lattice. We assume that the interface is in the plane (001). The
$\text{Ga}^{3+}$ ions form a square lattice with the
sides directed in [110] and [1\=10]. The ``electron" (wave packet) will then move isotropically with an effective
mass $m_{1}$. The $\text{As}^{3-}$ ions also form a square lattice at a different height in
[001]. The ``holes", each having a positive charge, will move similarly with an effective mass $m_{2}$.
A longitudinal phonon moving in [110] or in [1\=10] can generate a charge (current) density variations,
establishing an interaction between the phonon and the electron (phonon). If one phonon exchange is considered
between the electron and the fluxon, a second-order perturbation calculation establishes an effective
electron-fluxon interaction
\begin{equation}
|V_{q}V'_{q}| \frac{\hbar \omega_{q}}{(\epsilon_{|\mathbf{p}+\mathbf{q}|}-\epsilon_{p})^{2}
-(\hbar \omega_{q})^{2}}\,,
\end{equation}
where $\mathbf{q}$ $(\hbar\omega_{q})$ is the phonon momentum (energy), $V_{q}$ ($V'_{q}$) the interaction
strength between the electron (fluxon) and the phonon. If the energies
($\epsilon_{\mathbf{p}+\mathbf{q}},\,\epsilon_{p}$)
of the final and initial electron states are equal, the effective interaction is attractive.

Following Bardeen-Cooper-Schrieffer (BCS) [7], we start with a Hamiltonian $H$ with the phonon variables
eliminated:
\begin{eqnarray}
H&=&{\sum_{\bold{k}}}\sum_s \epsilon^{(1)}_k n^{(1)}_{\bold{k}s}
+{\sum_{\bold{k}}}\sum_s \epsilon^{(2)}_k n^{(2)}_{\bold{k}s}
+{\sum_{\bold{k}}}\sum_s \epsilon^{(3)}_k n^{(3)}_{\bold{k}s} \nonumber \\
&&-v_0 {\sum_{\bold{q}}}'{\sum_{\bold{k}}}'{\sum_{\bold{k'}}}'{\sum_s}
\left[ B^{(1)\dagger}_{\bold{k}'\bold{q}s} B^{(1)}_{\bold{k}\bold{q}s}
+ B^{(1)\dagger}_{\bold{k}'\bold{q}s} B^{(2)\dagger}_{\bold{k}\bold{q}s} \right. \nonumber \\
&& \qquad + \left. B^{(2)}_{\bold{k}'\bold{q}s} B^{(1)}_{\bold{k}\bold{q}s}
+ B^{(2)}_{\bold{k}'\bold{q}s} B^{(2)\dagger}_{\bold{k}\bold{q}s} \right],
\end{eqnarray}
where $n^{(j)}_{\bold{k}s}$ is the number operator for the ``electron"(1) [``hole" (2), fluxon (3)]
at momentum $\bold{k}$ and spin $s$ with the energy $\epsilon^{(j)}_{ks}$. We represent the ``electron"
(``hole") number $n^{(j)}_{\bold{k}s}$ by $c^{(j)\dagger}_{\bold{k}s}c^{(j)}_{\bold{k}s}$, where
$c$ $(c^\dagger)$ are annihilation (creation) operators satisfying the Fermi anticommutation rules:
$\{ c^{(i)}_{\bold{k}s},c^{(j)\dagger}_{\bold{k}'s} \} \equiv c^{(i)}_{\bold{k}s}c^{(j)\dagger}_{\bold{k}'s'}
+c^{(j)\dagger}_{\bold{k}'s'}c^{(i)}_{\bold{k}s}=\delta_{\bold{k},\bold{k}'}\delta_{s,s'}\delta_{i,j}$,
$\{ c^{(i)}_{\bold{k}s},c^{(j)}_{\bold{k}',s'} \}=0$. We represent the fluxon number $n^{3}_{\bold{k}s}$ by
$a^\dagger_{\bold{k}s}a_{\bold{k}s}$, with $a(a^\dagger)$, satisfying the anticommutation rules.
$B^{(1)\dagger}_{\bold{kq} \; s} \equiv c^{(1)\dagger}_{\bold{k}+\bold{q}/2 \; s}
a^{\dagger}_{-\bold{k}+\bold{q}/2 \; -s}$, 
$B^{(2)}_{\bold{kq} \; s} \equiv c^{(2)}_{\bold{k}+\bold{q}/2 \; s}
a_{-\bold{k}+\bold{q}/2 \; -s}$. The prime on the summation means the restriction:
$0<\epsilon^{(j)}_{ks}<\hbar\omega_D$, $\omega_D$ = Debye frequency.
If the fluxons are replaced by the conduction electrons (``electrons", ``holes") our Hamiltonian $H$ is
reduced to the original BCS Hamiltonian, Eq. (24) of ref. 7. The ``electron" and ``hole" are generated,
depending on the energy contour curvature sign [8]. For example only ``electrons" (``holes"), are
generated for a circular Fermi surface with the negative (positive) curvature whose inside (outside) is
filled with electrons. Since the phonon has no charge, the phonon exchange cannot change the net charge.
The pairing interaction terms in Eq. (2) conserve the charge. The term
$-v_0B^{(1)\dagger}_{\bold{k}'\bold{q}s}B^{(1)}_{\bold{kq}s}$, where
$v_0 \equiv |V_qV'_q|(\hbar\omega_0A)^{-1}$, $A$ = sample area, is the pairing strength, generates
the transition in the ``electron" states. Similary, the exchange of a phonon generates a transition
in the ``hole" states, represented by $-v_0 B^{(2)}_{\bold{k}'\bold{q}s} B^{(2)\dagger}_{\bold{kq}s}$.
The phonon exchange can also pair-create and pair-annihilate ``electron" (``hole")-fluxon composites,
represented by $-v_0 B^{(1)\dagger}_{\bold{k}'\bold{q}s} B^{(2)\dagger}_{\bold{kq}s}$,
$-v_0 B^{(2)}_{\bold{k}'\bold{q}s} B^{(1)}_{\bold{kq}s}$. At 0 K the system can have equal numbers of
$-(+)$ c-bosons, ``electrons" (``holes") composites, generated by
$-v_0 B^{(1)\dagger}_{\bold{k}'\bold{q}s} B^{(2)\dagger}_{\bold{kq}s}$.

The c-bosons, each with one fluxon, will be called the fundamental (f) c-bosons. Their energies $w_{q}^{(j)}$
are obtained from [8]
\begin{eqnarray}
w_{q}^{(j)} \Psi(\mathbf{k},\mathbf{q}) &=& \epsilon_{|\mathbf{k}+\mathbf{q}|}^{(j)}
\Psi(\mathbf{k},\mathbf{q}) \nonumber \\
&-& (2\pi\hbar)^{-2}v_{0}^{*} \int' d^{2}k' \, \Psi(\mathbf{k}',\mathbf{q}) \, ,
\end{eqnarray}
where $\Psi(\mathbf{k},\mathbf{q})$ is the reduced wavefunction for the fc-boson; we neglected the fluxon energy.
The $v_{0}^{*}$ denotes the strength after the ladder diagram binding, see below. For small $q$, we obtain
\begin{equation}
w_{q}^{(j)} = w_{0}+(2/\pi)v_{F}^{(j)}q, \quad w_{0}=\frac{-\hbar\omega_{D}}{\exp{(v_{0}^{*}D_{0})^{-1}}-1} \, ,
\end{equation}
where $v_{F}^{(j)} \equiv (2\epsilon_{F}/m_{j})^{1/2}$ is the Fermi velocity and $D_{0} \equiv D(\epsilon_{F})$
the density of states per spin. Note that the energy $w_{q}^{(j)}$ depends \textit{linearly} on the momentum $q$.

The system of free fc-bosons undergoes a Bose-Einstein condensation (BEC) in 2D at the critical temperature [8]
\begin{equation}
k_{B}T_{c}=1.24\;\hbar v_{F} n_{0}^{1/2} \, .
\end{equation}
The interboson distance $R_{0} \equiv n_{0}^{1/2}$ calculated from this expression is
$1.24\,\hbar v_{F}(k_{B}T_{c})^{-1}$. The boson size $r_{0}$ calculated from Eq. (4), using the uncertainty
relation $(q_{\text{max}}r_{0}\sim\hbar)$ and $|w_{0}|\sim k_{B}T_{c}$, is
$(2/\pi)\hbar v_{F}(k_{B}T_{c})^{-1}$,
which is a few times smaller than $R_{0}$. Hence, the bosons do not overlap in space, and the model of free bosons
is justified. For GaAs/AlGaAs, $m^{*}=0.067m_{c}$, $m_{c}=\text{electron mass}$. For the 2D electron
density $10^{11}$ cm$^{-2}$, we have $v_{F}=1.36\times 10^{6}$ cm s$^{-1}$. Not all electrons are bound with
fluxons since the simultaneous generations of $\pm$ fc-bosons is required. The minority carrier (``hole") density
controlls the fc-boson density. For $n_{0}=10^{10}$ cm$^{-2}$, $T_{c}=1.29$ K, which is reasonable.

In the presence of Bose condensate below $T_{c}$ the unfluxed electron carries the energy
$E_{k}^{(j)}=\sqrt{\epsilon_{k}^{(j)\,2}+\Delta^{2}}$, where the quasi-electron energy gap $\Delta$ is the
solution of
\begin{eqnarray}
1&=&v_{0}D_{0}\int_{0}^{\hbar\omega_{D}} d\epsilon \, \frac{1}{(\epsilon^{2}+\Delta^{2})^{1/2}} \nonumber \\
&& \times\left\{ 1+\exp\left[-\beta(\epsilon^{2}+\Delta^{2})^{1/2}\right] \right\}^{-1} \! ,
\;\; \beta\equiv\frac{1}{k_{B}T}\, . \nonumber \\
\end{eqnarray}
Note that the gap $\Delta$ depends on $T$. At $T_{c}$, there is no condensate and hence $\Delta$ vanishes.

Now the moving fc-boson below $T_{c}$ has the energy $\tilde{w}_{q}$ obtained from
\begin{eqnarray}
\tilde{w}_{q}^{(j)}\Psi(\mathbf{k},\mathbf{q})&=&E_{|\mathbf{k}+\mathbf{q}|}^{(j)}\Psi(\mathbf{k},\mathbf{q})
\nonumber \\ &&
-(2\pi\hbar)^{-2}v_{0}^{*}\int'd^{2}k'\Psi(\mathbf{k}',\mathbf{q}) \, ,
\end{eqnarray}
where $E^{(j)}$ replaced $\epsilon^{(j)}$ in Eq. (3). We obtain
\begin{equation}
\tilde{w}_{q}^{(j)}=\tilde{w}_{0}+(2/\pi)v_{F}^{(j)}q \equiv w_{0}+\epsilon_{g}+(2/\pi)v_{F}^{(j)}q \, ,
\end{equation}
where $\tilde{w}_{0}(T)$ is determined from
$1=D_{0}v_{0}\int_{0}^{\hbar\omega_{D}}d\epsilon\,[|\tilde{w}_{0}|+(\epsilon^{2}+\Delta^{2})^{1/2}]^{-1}$.
The energy difference: $\tilde{w}_{0}(T)-w_{0} \equiv \epsilon_{g}(T)$ represents the $T$-dependent
\textit{energy gap}. The energy $\tilde{w}_{g}$ is negative.
Otherwise, the fc-boson should break up. This limits $\epsilon_{g}(T)$ to be $|w_{0}|$ at 0 K. The
$\epsilon_{g}$ declines to zero as the temperature approaches $T_{c}$ from below.

The fc-boson, having the linear dispersion (8), can move in all directions in tha plane with the constant speed
$(2/\pi)v_{F}^{(j)}$. The supercurrent is generated by the $\pm$ fc-bosons condensed monochromatically at the
momentum directed along the sample length. The supercurrent density (magnitude) $J$, calculated by the rule:
$(\text{charge } e^{*}) \times (\text{carrier density } n_{0}) \times (\text{drift velocity } v_{d})$, is
\begin{equation}
J \equiv e^{*}n_{0}v_{d} = e^{*}n_{0}(2/\pi)|v_{F}^{(1)}-v_{F}^{(2)}| \, .
\end{equation}
The induced Hall field (magnitude) $E_{H}$ equals $v_{d}B$. The magnetic flux is quantized
$B=n_{\phi}(h/e)$, $n_{\phi}=\text{fluxon density}$. Hence we obtain
\begin{equation}
\rho_{H} \equiv \frac{E_{H}}{J} = \frac{v_{d}B}{e^{*}n_{0}v_{d}} = \frac{1}{e^{*}n_{0}}n_{\phi}
\left( \frac{h}{e} \right) \, .
\end{equation}
If $e^{*}=e$, $n_{\phi}=n_{0}$, we obtain $\rho_{H}=h/e^{2}$ in agreement with the plateau value observed.

The model can be extended to the integer QHE at $\nu = P$ $(Q=1)$. The field magnitude is less.
The LL degeneracy $(eBA/h)$ is linear in $B$, and hence the lowest $P$ LL's must be considered. The fc-boson
density $n_{0}$ per LL is the electron density $n_{e}$ over $P$ and the fluxon density $n_{\phi}$ is the boson
density $n_{0}$ over $P$:
\begin{equation}
n_{0}=n_{e}/P, \qquad n_{\phi}=n_{0}/P \, .
\end{equation}
At $\nu=1/2$ there are c-bosons, each with two fluxons. The c-fermions have a Fermi energy.
The $\pm$ c-fermions have effective masses. The Hall resistivity $\rho_{H}$ has a $B$-linear behavior while the
resistivity $\rho$ is finite.

Let us now take a general case $\nu=P/Q$, odd $Q$. Assume that there are $P$ sets of c-fermions with $Q-1$
fluxons, which occupy the lowest $P$ LL's. The c-fermions subject to the available $B$-field form c-bosons with
$Q$ fluxons. In this configuration the c-boson density $n_{0}$ and the fluxon density $n_{\phi}$ are given
by Eqs. (11). Using Eqs. (10) and (11) and assuming the fractional charge [9]
\begin{equation}
e^{*}=e/Q \, ,
\end{equation}
we obtain
\begin{equation}
\rho_{H} \equiv \frac{E_{H}}{J} = \frac{v_{d}}{e^{*}n_{0}v_{d}} n_{\phi} \left( \frac{h}{e} \right)
= \frac{Q}{P} \frac{h}{e^{2}} \, ,
\end{equation}
as observed. In our theory the integer $Q$ denotes the number of fluxons in the c-boson and the integer $P$ the
number of the LL's occupied by the parental c-fermions, each with $Q-1$ fluxons.

Our Hamiltonian in Eq. (2) can generate and stabilize the c-particles with an arbitrary number of fluxons.
For example a c-fermion with two fluxons is generated by two sets of the ladder diagram bindings, each between
the electron and the fluxon. The ladder diagram binding arises as follows. Consider a hydrogen atom. The
Hamiltonian contains kinetic energies of the electron and the proton, and the attractive Coulomb interaction. If
we regard the Coulomb interaction as a perturbation and use a purturbation theory, we can represent the
interaction proccess by an infinite set of ladder diagrams, each ladder step connecting the electron and the
proton. The energy eigenvalues of this system is not obtained by using the perturbation theory but they are
obtained by solving the Schr\"odinger equation directry. This example indicates that a two-body bound state is
represented by an infinite set of ladder diagrams and that the binding energy (the negative of the ground-state
energy) is calculated by a non-perturbative method.

Jain introduced the effective magnetic field [10]
\begin{equation}
B^{*} \equiv B-B_{\nu} = B - (1/\nu)n_{e}(h/e)
\end{equation}
relative to the standard field for the composite (c-) fermion at the even-denominator fraction. We extend this
to the bosonic (odd-denominator) fraction. This means that the c-particle moves field-free at the exact fraction.
The c-particle is viewd as the quasiparticle containing an electron circulating around $Q$ fluxons. The jumping
of the guiding centers (the CM of the c-particle) can occur as if they are subject to no B-field at the exact
fraction. The excess (or deficit) of the magnetic field is simply the effective magnetic field $B^{*}$. The
plateau in $\rho_{H}$ is formed due to the Meissner effect. Consider the case of zero temperature near $\nu=1$.
Only the energy $E$ matters. The fc-bosons are condensed with the ground-state energy $w_{0}$, and hence the
system energy $E$ at $\nu=1$ is $2N_{0}w_{0}$, where $N_{0}$ is the number of $-$ fc-bosons (or $+$ fc-bosons).
The factor 2 arises since there are $\pm$ fc-bosons. Away from $\nu=1$ we must add the magnetic field energy
$(2\mu_{0})^{-1}A(B^{*})^{2}$, so that
\begin{equation}
E=2N_{0}w_{0}+(2\mu_{0})^{-1}A(B^{*})^{2} \, .
\end{equation}
When the field is reduced, the system tries to keep the same number $N_{0}$ by sucking in the flux lines. Thus
the magnetic field becomes inhomogeneous outside the sample, generating the magnetic field energy
$(2\mu_{0})^{-1}A(B^{*})^{2}$. If the field is raised, the system tries to keep the same number $N_{0}$ by
expeling out the flux lines. The inhomogeneous fields outside raise the field energy as well. There is a critical
field $B_{c}^{*}=(4\mu_{0}|w_{0}|)^{1/2}$. Beyond this value, the superconducting state is destroyed, generating
a symmetric exponential rise in $R$. In our discussion of the Hall resistivity plateau we used the fact that the
ground-state energy $w_{0}$ of the fc-boson is negative, that is, the c-boson is bound. Only then the critical
field $B_{c}^{*}=(4\mu_{0}|w_{0}|)^{1/2}$ can be defined. Here the phonon exchange attraction played an important
role. The repulsive Coulomb interaction, which is the departure point of the prevalent theories [2,11], cannot
generate a bound state.

In the presence of the supercondensate
the non-condensed c-boson has an energy gap $\epsilon_g$.
Hence the noncondensed c-boson density has the activation energy type exponential
temperature-dependence:
\begin{equation}
\exp[- \epsilon_g/(k_BT)] \; .
\end{equation}
In the prevalent theories
the energy gap for the fractional QHE is identified as the sum of the creation energies of a quasi-electron
and a quasi-hole [11] With this view it is difficult to explain why the activation-energy type
temperature dependence shows up in the steady-state quantum transport.
Some authors argue that the energy gap $\epsilon_g$ for the integer QHE is due to the LL separation
= $\hbar \omega_0$.
But the separation $\hbar \omega_0$ is much greater than the observed $\epsilon_g$.
Besides from this view one cannot obtain the activation-type energy dependence.

The BEC occurs at each LL, and therefore the c-boson density $n_{0}$ is less for high $P$, see Eq. (11), and the
strengths become weaker as $P$ increases.

We are now ready to discuss the QHE under radiation. The experiments by Mani \textit{et al}. [4] indicate that the
applied radiation excites a large number of ``holes" in the system. Using these ``holes" and the preexisting
``electrons" the phonon exchange can pair-create $\pm$ c-bosons, which condense below $T_{c}$ in the
excited channel. The condensed c-bosons in motion are responsible for the supercurrent. In the presence of the
condensed c-bosons, the non-condensed c-bosons have an energy gap $\epsilon_{g}$, and therefore they are absent
at 0 K. The c-fermions in the base channel will have the energies $E_{p}=(\epsilon_{p}^{2}+\Delta^{2})^{2}$ but
their energy spectra have no gap. Hence they are not completely suppressed at the lowest temperatures. They
contribute a small normal current, and are responsible for the $B$-linear behavior of the Hall resistivity
observed.

Mani \textit{et al}. experiments [1], Fig. 2, show that the strength of the superconducting state
does not change much for the radiation frequency $\omega$ in the range (0.47, 110) GHz. This feature may come as
follows. The 2D density of states for the conduction electrons associated with the circular Fermi surface
is independent of the electron energy, and hence the number of the excited electrons is roughly independent of
the radiation energy (frequency). The ``hole"-like excitations are absent with no radiation. We suspect
that the ``hole"-band edge is a distance $\epsilon_{0}$ away from the system's Fermi level. This means that if the
radiation energy $\hbar \omega$ is less than $\epsilon_{0}$, the radiation can generate no superconducting state.
This feature can be checked by applying radiation of frequencies lower than 0.47 GHz.

In summary the QHE under radiation is the QHE at the upper channel. The condensed c-bosons generate
a superconducting state with a gap $\epsilon_{g}$ in the c-boson energy spectrum. The supercondensate changes the
c-fermion energy from $\epsilon_{k}$ to $(\epsilon_{k}^{2}+\Delta^{2})^{1/2}$ in the base channel. This energy
spectrum has no gap, and hence the c-fermions cannot be suppressed completely at the lowest temperatures, and
generate a finite resistive current accompanied by the Hall field. This explains the $B$-linear Hall resistivity.
Our microscopic theory can be tested experimentally by the detection of (a) the ``hole"-like excitations by a
circularly polarized laser, (b) the bosonic state at $\nu=4/3$ and $1/3$, (c) the ``hole" band edge.

We thank Dr. Mani for fruitful discussion.

\end{document}